\documentclass{article} 
\usepackage{iclr2023_workshop,times}

\usepackage{amsmath,amsfonts,bm}

\def\eqref#1{equation~\ref{#1}}

\def\1{\bm{1}}

\DeclareMathAlphabet{\mathsfit}{\encodingdefault}{\sfdefault}{m}{sl}
\SetMathAlphabet{\mathsfit}{bold}{\encodingdefault}{\sfdefault}{bx}{n}

\usepackage{hyperref}
\usepackage{url}
\usepackage{graphicx,kantlipsum,setspace}

\title{Invertible mapping between fields in CAMELS}

\makeatletter
\let\@fnsymbol\@arabic
\makeatother

\author{Sambatra Andrianomena \thanks{South African Radio Astronomy Observatory, Cape Town 7925, \texttt{andrianomena@gmail.com.}}~~\thanks{Department of Physics \& Astronomy University of the Western Cape, Cape Town 7535.}~~~~~~~~~~
Sultan Hassan \thanks{Center for Cosmology and Particle Physics, Department of physics, New York University, 726 Broadway, New York, NY, 10003, \texttt{shassan@flatironinstitute.org}.}~~\footnotemark[2]~~\footnotemark[4]~~~~~~~~~
Francisco Villaescusa-Navarro \thanks{Center for Computational Astrophysics, Flatiron Institute, New York, NY 10010, \texttt{villaescusa.francisco@gmail.com}.}
}

\iclrfinalcopy 
\begin{document}

\maketitle

\begin{abstract}
We build a bijective mapping between different physical fields from hydrodynamic CAMELS simulations. We train a CycleGAN on three different setups: translating dark matter to neutral hydrogen (Mcdm-HI), mapping between dark matter and magnetic fields magnitude (Mcdm-B), and finally predicting magnetic fields magnitude from neutral hydrogen (HI-B). We assess the performance of the models using various summary statistics, such as the probability distribution function (PDF) of the pixel values and 2D power spectrum ($P(k)$). Results suggest that in all setups, the model is capable of predicting the target field from the source field and vice versa, and the predicted maps exhibit statistical properties which are consistent with those of the target maps. This is indicated by the fact that the mean and standard deviation of the PDF of maps from the test set is in good agreement with those of the generated maps. The mean and variance of $P(k)$ of the real maps agree well with those of generated ones. The consistency tests on the model suggest that the source field can be recovered reasonably well by a forward mapping (source to target) followed by a backward mapping (target to source). This is demonstrated by the agreement between the statistical properties of the source images and those of the recovered ones.       
\end{abstract}

\section{Introduction}

The upcoming generation of surveys (e.g. SKA) will be able to map neutral hydrogen via HI intensity mapping \citep{santos2015cosmology}. This powerful cosmological probe will help us further our understanding of large-scale structure. To extract the relevant information about the matter field from these surveys, we usually resort to summary statistic, such as power spectrum. This is challenging in the non-linear regime due to the contamination of the signal by the baryonic physics and hence requires higher order statistics. The other approach is to carry out the analysis at the field level, e.g. inference or direct mapping between fields. Previous studies demonstrated the feasibility of building a mapping at the field level between baryons and dark matter \citep{wadekar2021hinet, villanueva2021removing, bernardini2022ember}. In light of those existing works, and by utilizing generative adversarial networks (CycleGAN), we aim at building a bijective map between dark matter and two observables, namely neutral hydrogen and magnetic field magnitude. 
This is crucial since with a single training, it is possible to either paint the dark matter from simulation with baryons or directly infer its distribution from maps of observables obtained from different surveys in the near future.







\section{Methods}
\label{methods}
\subsection{Data}
\label{data}

In this study, we use the publicly available CAMELS Multifields Dataset (CMD) \citep{CMD} which contains thousands of 2D field maps generated from state-of-the-art hydrodynamics simulations\citep{villaescusa2021camels}. We consider $256\times256$ pixels 2D maps, which cover an area of $25\times25~(h^{-1}{\rm Mpc})^2$ at $z=0$, of dark matter (Mcdm), neutral hydrogen (HI) and magnetic fields magnitude (B) from the IllustrisTNG LH set. The choice of the fields in this simple scenario is based on the aim of inferring the matter distribution from observables. In total, each field corresponds to 15000 2D images, each characterized by a set of 6 parameters: matter density ($\Omega_{8}$), the amplitude of matter power spectrum ($\sigma_{8}$), the stellar feedbacks ($A_{\rm SN1}$, $A_{\rm SN1}$) and AGN feedbacks ($A_{\rm AGN1}$, $A_{\rm AGN2}$). 

\subsection{Model and training}
\label{model}
To build an invertible mapping between two different fields, we make use of CycleGAN \citep{zhu2017unpaired}, an improvement on ``pix2pix'' method \citep{isola2017image} which is trained on paired examples to achieve image-to-image translation. The approach consists of building a function $\mathcal{G}: X\rightarrow Y$ that maps a source field $X$ to a target field $Y$ (forward mapping), simultaneously with another function $\mathcal{F}: Y\rightarrow X$ that translates the target to the source field. To this end, two generators $G_{X}$ (representing $\mathcal{F}$ and producing the source field images) and $G_{Y}$ (representing $\mathcal{G}$ and producing the target field images) are trained with two adversarial discriminators $D_{X}$ and $D_{Y}$ respectively. Following the prescription in \citet{zhu2017unpaired}, there are two main components to the loss function for the training. The adversarial loss -- for each  of the pair ($G_{X}, D_{X}$) and ($G_{Y}, D_{Y}$) -- is given by \citep{zhu2017unpaired}
\begin{equation}
    \mathcal{L}_{\rm GAN}(G_{X}, D_{X}, Y, X) = \mathbb{E}_{x\sim p_{\rm data}(x)}[{\rm log}D_{X}(x)] + \mathbb{E}_{y\sim p_{\rm data}(y)}[{\rm log}(1 - D_{X}(G_{X}(y)))]
\end{equation}
and 
\begin{equation}
    \mathcal{L}_{\rm GAN}(G_{Y}, D_{Y}, X, Y) = \mathbb{E}_{y\sim p_{\rm data}(y)}[{\rm log}D_{Y}(y)] + \mathbb{E}_{x\sim p_{\rm data}(x)}[{\rm log}(1 - D_{Y}(G_{Y}(x)))]
\end{equation}
where $x\sim p_{\rm data}(x)$ and $y\sim p_{\rm data}(y)$ are the data distributions of the source images ($x\in X$) and target images ($y\in Y$) respectively. The consistency loss ensures that the functions $\mathcal{G}$ and $\mathcal{F}$ are inverse of each other such that $G_{X}(G_{Y}(x))\approx x$ and  $G_{Y}(G_{X}(y))\approx y$. In other words, an input $x$ (or $y$) is recovered by applying $G_{X}$ on $G_{Y}(x)$ (or applying $G_{Y}$ on $G_{X}(y)$).
\begin{figure}[h]
\begin{center}
\includegraphics[width=0.65\textwidth]{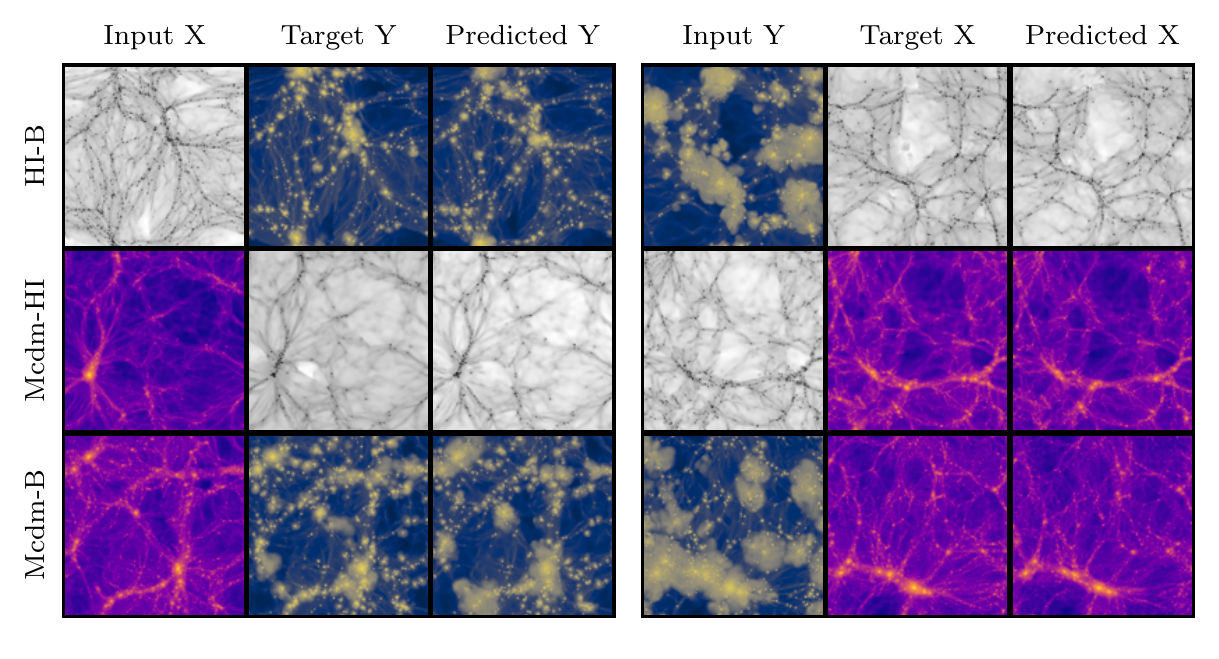}
\end{center}
\caption{The top, middle, and bottom rows show the results from mapping HI-B, Mcdm-HI, and Mcdm-B respectively. 
It is worth noting that in a X-Y setting, X and Y designate the source and target fields, respectively.}\label{fig:maps-results}
\end{figure}
The consistency loss is given by \citep{zhu2017unpaired}
\begin{equation}
    \mathcal{L}_{\rm cycle} =  \mathbb{E}_{x\sim p_{\rm data}(x)}[||G_{X}(G_{Y}(x)) - x||_{1}] + \mathbb{E}_{y\sim p_{\rm data}(y)}[||G_{Y}(G_{X}(y)) - y||_{1}],
\end{equation}
where $||.||_{1}$ denotes the mean absolute error (L1Loss). To further enforce a unique prediction of a given input such that $G_{Y}(y) \approx y$ and $G_{X}(x) \approx x$, an identity loss 
\begin{equation}
    \mathcal{L}_{\rm id} = \mathbb{E}_{y\sim p_{\rm data}(y)}[||G_{Y}(y) - y||_{1}] + \mathbb{E}_{x\sim p_{\rm data}(x)}[||G_{X}(x) - x||_{1}]
\end{equation}
is used. The total loss is then given by 
\begin{equation}
    \mathcal{L}_{\rm tot} = \mathcal{L}_{\rm GAN}(G_{X}, D_{X}, Y, X) + \mathcal{L}_{\rm GAN}(G_{X}, D_{X}, Y, X) + \lambda_{\rm cycle}\mathcal{L}_{\rm cycle} + \lambda_{\rm id}\mathcal{L}_{\rm id},
\end{equation}
where both $\lambda_{\rm cycle}$ and $\lambda_{\rm id}$ are constants that characterize the contributions of the consistency cycle loss and identity loss respectively. Based on the prescription in \citet{zhu2017unpaired}, we have that $\lambda_{\rm cycle} = 10$ and $\lambda_{\rm id} = 5$ during training. \\
The generators $G_{Y}$ and $G_{X}$, which have the same architecture, comprise a stage that downsamples the inputs (similar to encoding), 9 residual layers mimicking a bottleneck in the variational encoder and finally, a stage that upsamples the output from the bottleneck (similar to decoding). The discriminators $D_{Y}$ and $D_{X}$, which are also identical, consist of chaining up 5 convolutional layers where the first three downsample the input by using stride = 2. To have a bit more control on the topology of the generated maps, we condition the input of each component of the model ($G_{X}$, $D_{X}$, $G_{Y}$ and $D_{Y}$) on the underlying cosmology and astrophysics, i.e the parameters $\Omega_{8}$, $\sigma_{8}$, $A_{\rm SN1}$, $A_{\rm SN1}$, $A_{\rm AGN1}$ and $A_{\rm AGN2}$. The array of parameters of shape $1\times 6$ is passed through a dense layer with 4096 units whose output is reshaped to $64\times64$, upsampled to $256\times256$ via interpolation, and finally concatenated along the channel to the input image. The model is trained for 100 epochs with 12000 unpaired examples, i.e. shuffling the set of source images such that they don't match the target images, using Adam optimizer with a learning rate of 0.0002. By setting the batch size to 1, following \citet{zhu2017unpaired}, each epoch takes about 109 minutes on a NVIDIA GeForce GTX 1080 Ti.





\section{Results}
\label{headings}
We present in Figure~\ref{fig:maps-results} some predictions by the generators ($G_{Y}$ and $G_{X}$) using input images from test set. 
\vspace{-0.2cm}
\begin{figure}[h]
\begin{center}
\includegraphics[width=0.58\textwidth]{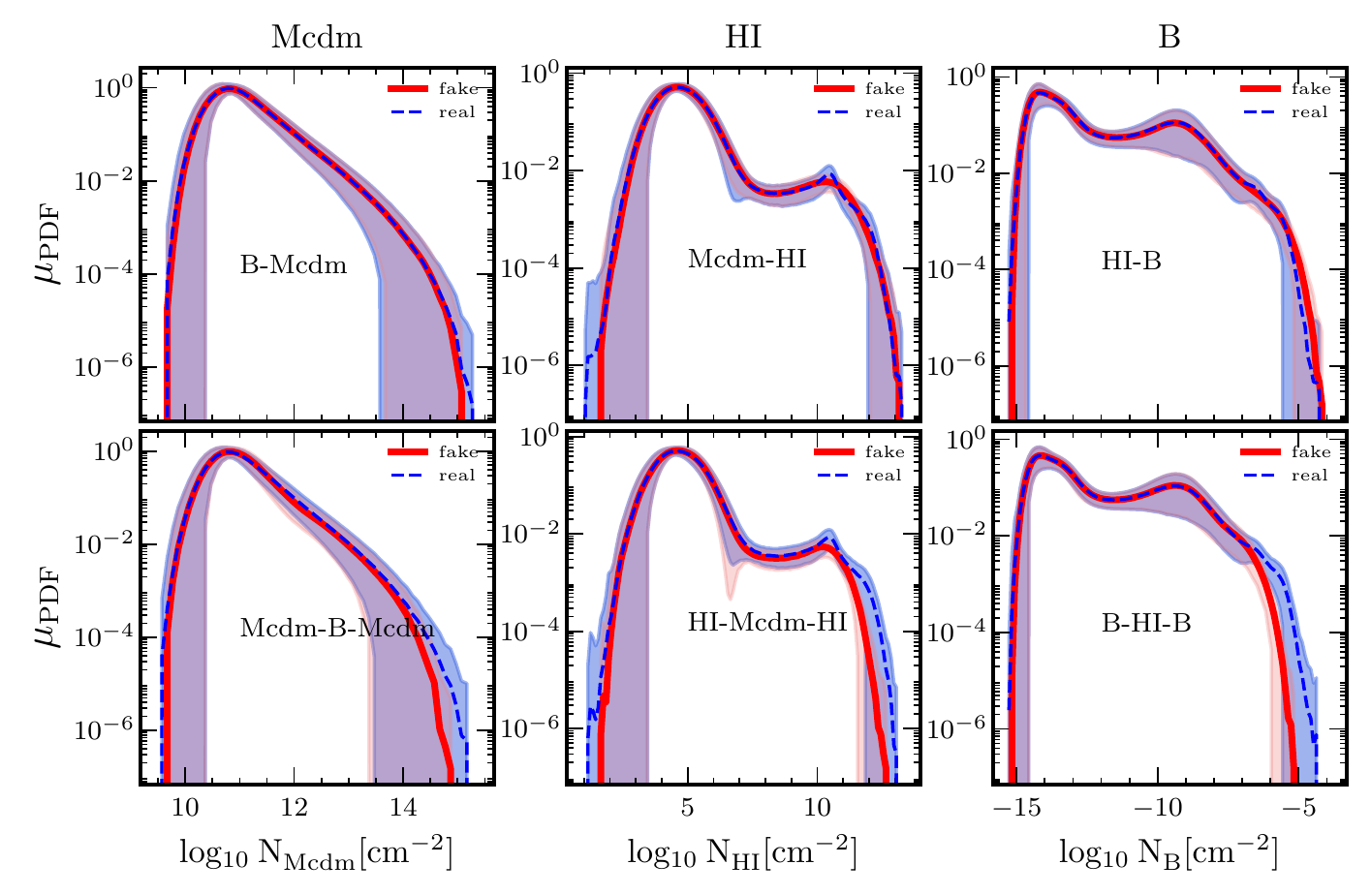}
\end{center}
\caption{Comparing probability distribution function (PDF) of pixel values of the simulated and generated maps. Results corresponding to each field are shown in each column. Solid red and dashed blue lines denote PDFs of fake and real maps, respectively. Whereas red and blue shaded areas correspond to the standard deviations of the PDFs of fake and real maps, respectively. Each column corresponds to PDF of fields in different setups. The top rows show the comparison between the PDFs of predicted and true maps of a field in each setup. The bottom rows show the consistency test.
}\label{fig:pdf}
\end{figure}
\vspace{-0.1cm}
Each column in Figure~\ref{fig:maps-results} corresponds to the input, target and prediction in a given setup, e.g. HI-B.  In the first three columns of each row, we show the result from the forward mapping, i.e. the input is the source field X and the output is the target field Y. The last three columns of each row show the results related to the backward mapping, i.e. the input is the target field Y which is translated to the source field X. Visually, the output of the map by the generators are in good agreement with the inputs and the quality is comparable to that of the data from IllustrisTNG, overall. 
\begin{figure}[h]
\begin{center}
\includegraphics[width=0.58\textwidth]{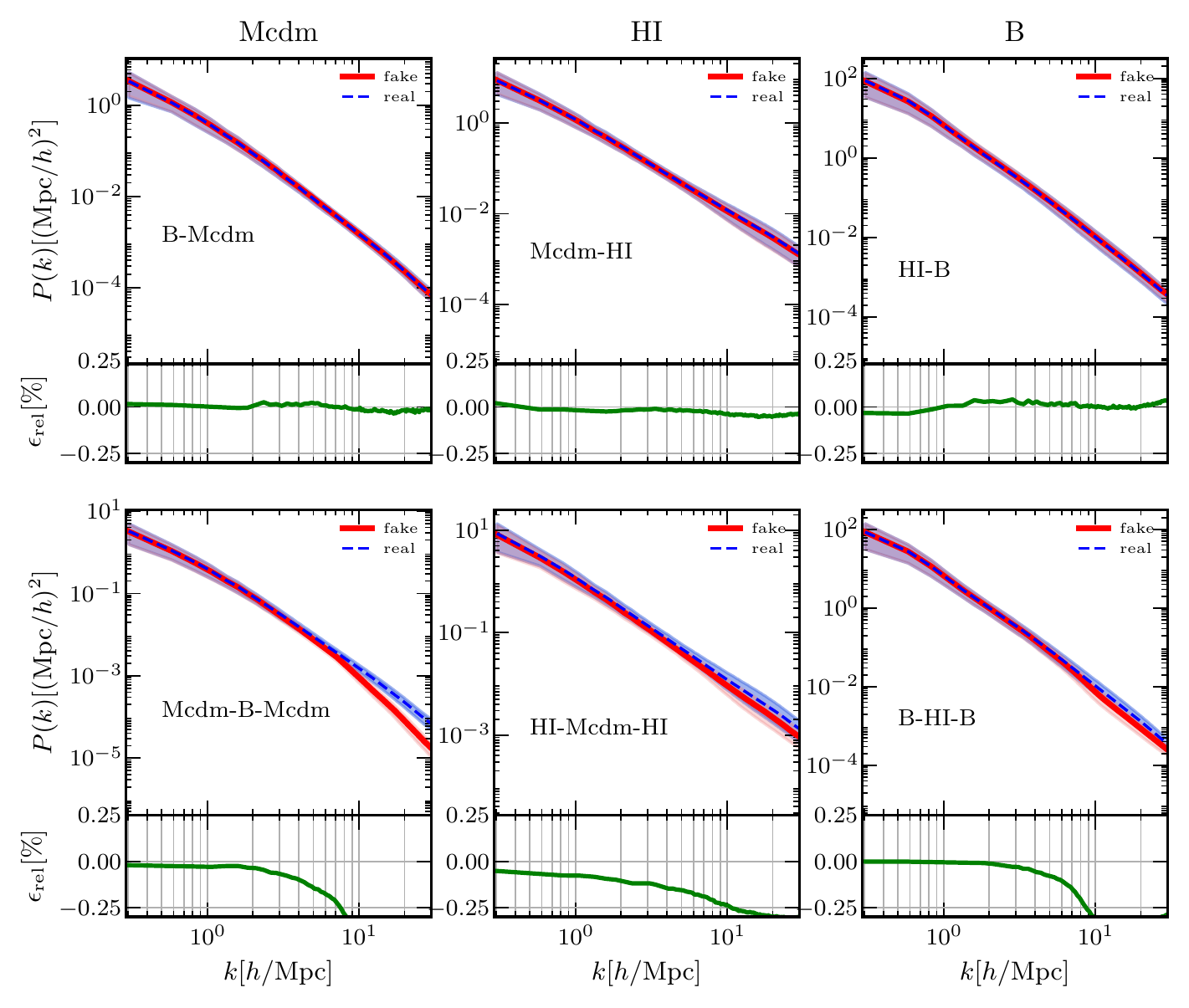}
\end{center}
\caption{Comparing the resulting power spectra of the simulated and generated maps. Results corresponding to each field are shown in each column. Solid red and dashed blue lines denote the averaged power spectrum ($P(k)$) of fake and real maps, respectively. Whereas red and blue shaded areas correspond to the standard deviations of the $P(k)$ of fake and real maps, respectively. The relative error between the two power spectra for each field in each setup is indicated by the solid green line.}\label{fig:power-spectra}
\end{figure}
However it appears that predicting the magnetic field B from dark matter or neutral hydrogen seems to be a bit more challenging, as evidenced by the more noticeable difference in the map features between the ground truth and prediction (see 2nd and 3rd columns of both top and bottom rows).\\ 
The first metric we use to assess how well the model performs is the probability distribution function (PDF) of pixel intensities. The test set for each setup comprises of 1000 images unseen by the model during training. We then compute the mean $\mu_{\rm PDF}$ and standard deviation $\sigma_{\rm PDF}$ of PDF of each field in the test set in each setup. In Figure~\ref{fig:pdf}, we present the PDFs of each field in some setups in each column. The top rows show the results from either forward or backward mappings, whereas the results of the consistency test -- assessing if an input is recovered by applying forward and backward mappings sequentially -- is presented at the bottom row. Results suggest that the model is capable of predicting the source and target fields in each setup using $G_{X}$ and $G_{Y}$ respectively, as evidenced by both $\mu_{\rm PDF}$ and $\sigma_{\rm PDF}$  of the data in good agreement with those of the generated maps of each field. We also find that in general, the bijective mapping is achieved to a good accuracy, i.e. $G_{X}(G_{Y}(x))\approx x$ and  $G_{Y}(G_{X}(y))\approx y$, as demonstrated by the $\mu_{\rm PDF}$ and $\sigma_{\rm PDF}$  of the data which agree well with those of the recovered outputs. The relatively small discrepancy at the high end of the distributions can be accounted for by the small number of pixels having those values in the training data, i.e. overdense regions are rare.
The other metric used in our investigation is the auto-power spectrum $P(k)$.
Similar to the results presented in Figure~\ref{fig:pdf}, we show the $P(k)$ of the images produced by the forward/backward mapping and the consistency test at the top and bottom rows of Figure~\ref{fig:power-spectra} respectively. Each panel shows the mean and standard deviation of the $P(k)$ of both real (dashed blue) and fake maps (solid red), and the solid green line at the bottom of each panel shows the relative difference $\left(\frac{P_{\rm fake}(k)}{P_{\rm real}(k)} - 1\right)$ between the two $P(k)'$s (fake and real maps). It is clear that the forward and backward mappings are able to produce maps with clustering properties in good agreement with those of the data. Moreover, the mean and variance of $P(k)$ of the recovered maps from the consistency check agree reasonably well with those of the maps from IllustrisTNG, regardless of the increasing difference on larger scales $k > 10\>h/{\rm Mpc}$. Our result for Mcdm-HI (see Figure~\ref{fig:power-spectra} top middle panel) is comparable to what was obtained by \cite{wadekar2021hinet} where a standard UNet architecture was used to convert dark matter density to HI maps.






\section{Conclusion}
\label{conclusion}
We have made use of CycleGAN model to build a bijective model that can map different physical fields from 2D maps created from the CAMELS state-of-the-art hydrodynamic simulations. 
Results show that the predicted maps exhibit statistical properties that agree well with those from the dataset used for training. Moreover, the condition for bijective mapping is met, as demonstrated by the consistency test. By applying the composition function $\mathcal{F}\circ\mathcal{G}(x)$ (or $\mathcal{G}\circ\mathcal{F}(y)$), an input map $x$ (or $y$) is recovered reasonably while the statistics being preserved. This work represents a step forward towards establishing an efficient {\it direct} mapping between different observables, and hence maximizing the scientific return of future multi-wavelength surveys.

\subsubsection*{Acknowledgments}
SA acknowledges financial support from the South African Radio Astronomy Observatory (SARAO). FVN and SH acknowledge support provided by the Simons Foundation. SH also acknowledges support for Program number HST-HF2-51507 provided by NASA through a grant from the Space Telescope Science Institute, which is operated by the Association of Universities for Research in Astronomy, incorporated, under NASA contract NAS5-26555. The CAMELS project is supported by NSF grant AST 2108078.

\bibliography{iclr2023_workshop}

\begin{thebibliography}{8}
\providecommand{\natexlab}[1]{#1}
\providecommand{\url}[1]{\texttt{#1}}
\expandafter\ifx\csname urlstyle\endcsname\relax
  \providecommand{\doi}[1]{doi: #1}\else
  \providecommand{\doi}{doi: \begingroup \urlstyle{rm}\Url}\fi

\bibitem[Bernardini et~al.(2022)Bernardini, Feldmann, Angl{\'e}s-Alc{\'a}zar,
  Boylan-Kolchin, Bullock, Mayer, and Stadel]{bernardini2022ember}
Mauro Bernardini, Robert Feldmann, Daniel Angl{\'e}s-Alc{\'a}zar, Mike
  Boylan-Kolchin, James Bullock, Lucio Mayer, and Joachim Stadel.
\newblock From ember to fire: predicting high resolution baryon fields from
  dark matter simulations with deep learning.
\newblock \emph{Monthly Notices of the Royal Astronomical Society},
  509\penalty0 (1):\penalty0 1323--1341, 2022.

\bibitem[Isola et~al.(2017)Isola, Zhu, Zhou, and Efros]{isola2017image}
Phillip Isola, Jun-Yan Zhu, Tinghui Zhou, and Alexei~A Efros.
\newblock Image-to-image translation with conditional adversarial networks.
\newblock In \emph{Proceedings of the IEEE conference on computer vision and
  pattern recognition}, pp.\  1125--1134, 2017.

\bibitem[Santos et~al.(2015)Santos, Bull, Alonso, Camera, Ferreira, Bernardi,
  Maartens, Viel, Villaescusa-Navarro, Abdalla, et~al.]{santos2015cosmology}
Mario~G Santos, Philip Bull, David Alonso, Stefano Camera, Pedro~G Ferreira,
  Gianni Bernardi, Roy Maartens, Matteo Viel, Francisco Villaescusa-Navarro,
  Filipe~B Abdalla, et~al.
\newblock Cosmology with a ska hi intensity mapping survey.
\newblock \emph{arXiv:1501.03989}, 2015.

\bibitem[Villaescusa-Navarro et~al.(2021)Villaescusa-Navarro,
  Angl{\'e}s-Alc{\'a}zar, Genel, Spergel, Somerville, Dave, Pillepich,
  Hernquist, Nelson, Torrey, et~al.]{villaescusa2021camels}
Francisco Villaescusa-Navarro, Daniel Angl{\'e}s-Alc{\'a}zar, Shy Genel,
  David~N Spergel, Rachel~S Somerville, Romeel Dave, Annalisa Pillepich, Lars
  Hernquist, Dylan Nelson, Paul Torrey, et~al.
\newblock The camels project: Cosmology and astrophysics with machine-learning
  simulations.
\newblock \emph{The Astrophysical Journal}, 915\penalty0 (1):\penalty0 71,
  2021.

\bibitem[Villaescusa-Navarro et~al.(2022)Villaescusa-Navarro, Genel,
  Angles-Alcazar, Thiele, Dave, Narayanan, Nicola, Li, Villanueva-Domingo,
  Wandelt, et~al.]{CMD}
Francisco Villaescusa-Navarro, Shy Genel, Daniel Angles-Alcazar, Leander
  Thiele, Romeel Dave, Desika Narayanan, Andrina Nicola, Yin Li, Pablo
  Villanueva-Domingo, Benjamin Wandelt, et~al.
\newblock The camels multifield data set: Learning the universe’s fundamental
  parameters with artificial intelligence.
\newblock \emph{The Astrophysical Journal Supplement Series}, 259\penalty0
  (2):\penalty0 61, 2022.

\bibitem[Villanueva-Domingo \& Villaescusa-Navarro(2021)Villanueva-Domingo and
  Villaescusa-Navarro]{villanueva2021removing}
Pablo Villanueva-Domingo and Francisco Villaescusa-Navarro.
\newblock Removing astrophysics in 21 cm maps with neural networks.
\newblock \emph{The Astrophysical Journal}, 907\penalty0 (1):\penalty0 44,
  2021.

\bibitem[Wadekar et~al.(2021)Wadekar, Villaescusa-Navarro, Ho, and
  Perreault-Levasseur]{wadekar2021hinet}
Digvijay Wadekar, Francisco Villaescusa-Navarro, Shirley Ho, and Laurence
  Perreault-Levasseur.
\newblock Hinet: Generating neutral hydrogen from dark matter with neural
  networks.
\newblock \emph{The Astrophysical Journal}, 916\penalty0 (1):\penalty0 42,
  2021.

\bibitem[Zhu et~al.(2017)Zhu, Park, Isola, and Efros]{zhu2017unpaired}
Jun-Yan Zhu, Taesung Park, Phillip Isola, and Alexei~A Efros.
\newblock Unpaired image-to-image translation using cycle-consistent
  adversarial networks.
\newblock In \emph{Proceedings of the IEEE international conference on computer
  vision}, pp.\  2223--2232, 2017.

\end{thebibliography}
\bibliographystyle{iclr2023_workshop}


\end{document}